\documentclass[12pt,preprint]{aastex}
\usepackage{amsmath}

\shorttitle{Main-Belt Comet P/2012 F5 (Gibbs)}
\shortauthors{Moreno et al.}

\begin{document}


\title{A short-duration event as the cause of dust ejection from Main-Belt 
Comet P/2012 F5 (Gibbs)}


\author{F. Moreno\affil{Instituto de Astrof\'\i sica de Andaluc\'\i a, CSIC,
  Glorieta de la Astronom\'\i a s/n, 18008 Granada, Spain}
\email{fernando@iaa.es}}

\author{
J. Licandro\affil{Instituto de Astrof\'\i sica de Canarias,
  c/V\'{\i}a 
L\'actea s/n, 38200 La Laguna, Tenerife, Spain, 
\and 
 Departamento de Astrof\'{\i}sica, Universidad de
  La Laguna (ULL), E-38205 La Laguna, Tenerife, Spain}}  

\and

\author{
A. Cabrera-Lavers\affil{Instituto de Astrof\'\i sica de Canarias,
  c/V\'{\i}a 
L\'actea s/n, 38200 La Laguna, Tenerife, Spain, 
\and 
 Departamento de Astrof\'{\i}sica, Universidad de
  La Laguna (ULL), E-38205 La Laguna, Tenerife, Spain, 
\and 
GTC Project, E-38205 La Laguna, Tenerife, Spain}}


\begin{abstract}

We present observations and an interpretative model of the dust 
environment of Main-Belt Comet P/2010 
F5 (Gibbs).  The narrow dust trails observed can be interpreted
unequivocally as an impulsive event that took place around 2011 July 1st 
with an uncertainty of $\pm$10 days, and a duration of less than a
day, possibly of the order of a few hours. The 
best Monte Carlo dust model fits to the observed trail brightness 
imply ejection velocities in the range 8-10 cm
s$^{-1}$ for particle sizes between 30 cm and 130 $\mu$m. This weak
dependence of velocity on size contrasts with that expected from ice
sublimation, and agrees with that found recently 
for (596) Scheila, a likely impacted asteroid. The particles seen in the trail
are found to follow a power-law size distribution 
of index $\approx$--3.7. Assuming that
the slowest particles were ejected at the 
escape velocity of the nucleus, its size is constrained to about
200-300 m in diameter. The total ejected dust mass 
is $\gtrsim 5\times 10^8$ kg, with represents approximately 
4 to 20\% of the nucleus mass. 
\end{abstract}


\section{Introduction}

Main-Belt Comet P/2012 F5 (Gibbs) (hereafter P/Gibbs) 
was discovered in the course of the Mt. Lemmon Survey on UT 2012
Mar. 22.89 \citep{Gibbs12}. The object showed a narrow dust tail 7$\arcmin$ in
length, and was identified as a Main-Belt Comet (MBC) owing to its orbital
parameters.  MBCs have 
dynamical properties of asteroids (i.e.,
Tisserand parameters respect to Jupiter larger than 3), but
physical characteristics of comets (gas and/or dust emission). The
total members of this new class of Small Solar System Objects amount
to nine (the 9th becoming P/Gibbs), so that the statistics is still
poor so as to identify their global physical properties or 
dynamical history. For reviews on those objects, we refer to
\cite{Bertini11} and \cite{Jewitt12}. Most of the MBCs  
have been shown to be dynamically stable on timescales
of 100 Myr or longer \citep{IpatovHahn99,Hsieh12a,Hsieh12b,Hsieh12c,Stevenson12} suggesting that they are native
members of the Main Asteroid Belt and not captured objects from
elsewhere \citep{Hsieh09}. This is strongly supported by dynamical and 
spectroscopic arguments that show that some of them belong to well established 
asteroid collisional families and that their spectra is different to that of 
``normal'' comets \citep[e.g.][]{Licandro11}.   
However, there are some
members as 238P/Read and P/2008 R1 that are stable for 20-30 Myr
only \citep{Haghighipour09,Jewitt09}. Regarding their activity, some are clearly 
recurrent, as 133P/Elst-Pizarro and 238P
\citep{Hsieh10,Hsieh11}, while some others like
(596) Scheila appear to have ejected dust via an impulsive event that
might be associated to a collision
\citep{Jewitt11,YangHsieh11,Bodewits11,Moreno11a,Ishiguro11}. On the
other hand, 
P/2010 R2 (La Sagra) and 2006 VW139 appeared to be continuously active during
periods of at least 200 days and 100 days, respectively
\citep{Moreno11b, Licandro12,Hsieh12b}.

Observations and analysis of P/Gibbs have been recently shown by 
\cite{Stevenson12}, who report on the impulsive character of the 
emission of the dust, the outburst date, the mass of the ejecta, and establish 
limits to the nucleus size. In this paper we report images in 
the red spectral domain acquired with
instrumentation attached to the 10.4-m Gran Telescopio Canarias (GTC) 
of P/Gibbs at two different dates, and perform an interpretation of the
observed trail brightness using a forward Monte Carlo tail model, in order
to provide estimates of the dust emission times, the ejected mass, the particle
ejection velocities, and their size distribution function. We also set limits 
to the nucleus size based on the escape velocity, and compare our results to 
those by \cite{Stevenson12}.

\section{Observations and data reduction}

CCD images of P/Gibbs were collected under photometric conditions on
the nights of 18 May and 8 June 2012, using a Sloan $r^\prime$ filter in the
Optical System for Image and Low Resolution Integrated Spectroscopy
(OSIRIS) camera-spectrograph \citep{Cepa00,Cepa10} at the GTC. The
OSIRIS instrument consists of two Marconi CCD detectors, each with
2048$\times$4096 pixels and a total unvignetted field of view of
7.8$\arcmin\times$7.8$\arcmin$. The plate scale was 0.127
$\arcsec$/px, but we used a binning of 2$\times$2 pixels in order to
improve the signal to noise ratio, so that the spatial resolution of the
images is 485 km px$^{-1}$ and 540 km px$^{-1}$ on May 18.9 and June 8.9,
respectively. The images were bias subtracted and
flat-field corrected using
standard procedures. A total of 18 images were acquired each night and 
calibrated using standard stars. The images were 
converted to solar disk intensity units appropriate for the 
analysis in terms of dust tail models, and a  median stack image 
was obtained for analysis for each date (see Figure 1). As a result of
both the flux calibration and the median stacking procedure, we estimated a
total flux uncertainty in the combined images of 0.3 mag. The images
were finally rotated to the $(N,M)$ coordinate system \citep{Finson68}. 

\section{The Model}

We have performed an analysis of the two images shown in Figure 1 by 
a direct Monte Carlo dust tail model. This model was used to
characterize the dust environments of various comets and MBCs  
\citep[e.g.,][]{Moreno09,Moreno10,Moreno11b}, being applied in
particular to  
comet 67P/Churyumov-Gerasimenko, the target of {\it Rosetta} Mission
to arrive to the comet in 2014 \citep[the so-called 
Granada model, see][]{Fulle10}. The code is used to compute 
the trajectory of a large number of grains ejected from a 
cometary or asteroidal surface. We assume that the only governing 
forces on those dust particles are the 
solar gravity and the radiation pressure. Thus, the gravity
of the object itself is neglected, which constitutes a good approximation for
small-sized nuclei. In the case of cometary activity, we consider that
the particles are  
accelerated by gas drag from ice sublimation to their 
terminal velocities, which are the input ejection  
velocities considered in the model, that is applicable for any other
ejection mechanism. Once 
ejected, the particles describe a Keplerian
trajectory around the Sun, whose orbital elements are computed from
the terminal velocity and the ratio of the 
force exerted by the solar radiation pressure and the solar gravity
\citep[the $\beta$ parameter, see][]{Fulle89}. This parameter can be expressed as $\beta =
C_{pr}Q_{pr}/(2\rho r)$, where $C_{pr}$=1.19$\times$ 10$^{-3}$ kg
m$^{-2}$, $Q_{pr}$ is the radiation presure coefficient, 
and $\rho$ is the particle density. For each observation date, the trajectories
of a large number of dust particles are computed, and then their
positions on the $(N,M)$ plane are calculated. Finally, their contribution to
the tail brightness is computed. 

Before applying the model, we performed a preliminary analysis of the
images in terms of a synchrone map \citep{Finson68}, which provides us
with an
approximate idea on the time interval at which the particles were
ejected from the nucleus related to the observation date. Figure 2 
shows the two images as well as the corresponding synchrones at times
of $\pm$100, $\pm$20, $\pm$10 and 0 days relative to the 2011 July 3rd 
synchrone, the one which is approximately best aligned with the narrow 
trail in each image. The smallest sized particles in the 2011 July 3rd 
synchrones have $r\sim$130 $\mu$m and $r\sim$180 $\mu$m for the two observing
dates, respectively. These diagrams indicate that all the dust
particles having $r\gtrsim$130 $\mu$m  were ejected during 
a very short time interval around that date, because otherwise the 
trails would have been wider towards the lowermost portions of the
trails, even if they were ejected with zero velocity respect to the
nucleus. Therefore, an impulsive event must be the responsible of the
ejection of dust, so that an outburst, a collision with another body,
or a rotational disruption could be in principle invoked among the
responsible mechanisms. The precise event date, and its duration, is to 
be determined on the basis of the Monte Carlo dust tail analysis 
(see section 4).

An important aspect of the observations and the modeling is the
location of the asteroid nucleus. We will work under the assumption 
that the object nucleus is immersed in the dust cloud, and that
the optocenter of this cloud corresponds to the actual location of the
nucleus. We further assume that the observed brightness is dominated
by the dust, the contribution of the nucleus being negligible. This
hypothesis would be tested on the basis of the nucleus size derived from
arguments based on the escape velocity. 

A number of simplifying assumptions on the physical parameters must be made in 
order to make the problem tractable. Thus, we assume that the trail is 
composed of spherical particles of carbonaceous composition, having a 
refractive index at red wavelengths of m=1.88+0.71$i$ \citep{Edoh83}, 
which implies a geometric albedo of $p_v$=0.036, and a pressure
radiation coefficient of $Q_{pr}\sim$ 1 for particles of radius 
$r \gtrsim$1 $\mu$m \citep[][their Figure 5]{Moreno12}. 
We further assume that the particles 
have a density of 1000 kg m$^{-3}$. The particle 
size distribution is assumed to follow a power-law of index $\alpha$, and 
the terminal velocities are described as a function of the $\beta$ 
parameter as $v(\beta)=v_0\beta^{\gamma}$. This relationship is
generally accepted for the terminal velocities of comet dust, and also for
fragments ejected from collision experiments
\citep[e.g.,][]{Giblin98,Onose04}. 
The total ejected dust 
mass ($M_e$) and the maximum size of the ejected particles ($r_{max}$)
must be also specified, as well as the starting time ($t_s$) and the 
duration of the event ($\Delta t$). In summary, there are a total of
seven adjustable parameters to fit the observations ($\alpha$, $v_0$, 
$\gamma$, $M_e$, $r_{max}$, $t_s$, and $\Delta t$). 

\section{Results}

We attempted to fit the observed dust trail brightness 
by minimizing the function $\sigma=\sigma_1+\sigma_2$, where the
subscripts 1 and 2 correspond to the images obtained in 2012 May 18.9
and 2012 June 8.9, respectively, and   
$ \sigma_i=\sqrt{(\sum(I_{obs}(i)-I_{fit}(i))^2/N(i))}$, where
$I_{obs}(i)$ are the observed trail intensities and $I_{fit}(i)$ are the
fitted intensities, the summation being extended in principle to all the image
pixels $N(i)$. Since there are some regions in the images that are strongly
contaminated by bright field stars, we restricted the summation to
pixels outside those regions, and located mostly along the trails. 
The minimization procedure was performed by the multidimensional downhill 
simplex algorithm \citep{Nelder65}, using the FORTRAN implementation
described in \cite{Press92}. Each of those parameters influence the
derived trail brightness in different ways. Thus, $v_0$ controls the width of
the trail, $\gamma$ influences the variation of the width of the trail
along it, $\alpha$ constraints the slope of the 
brightness along the trail, the total ejected mass influences the
overall brightness, and the maximum particle size controls the
brightness mainly at the head of the trail, so that if $r_{max }$ is set to a
small value (say, smaller than several centimeters) the peak of 
brightness is displaced significantly trailward and the fit becomes
impossible. Regarding the minimum particle size, it is not a free
parameter, as it is found by the intersection of the synchrone that is
best aligned with the trails and the $N$ axis, which corresponds in our images 
to $\sim$130 $\mu$m and $\sim$180 $\mu$m for 2012 May 18.9 and 2012 June 8.9, 
respectively.

Since the downhill simplex method searches for a local minimum of the
function in the parameter space, we performed several runs by varying
the starting simplex. We found that all runs tend to converge to close
local minima, giving similar values of $\sigma$. The deepest global 
minimum was then taken as the best fit, which has
$\sigma$=4.9$\times$10$^{-15}$ solar disk units (see Table 1, and Figure 3). 
Unfortunately, this 
technique does not provide any estimate of the uncertainty in the 
derived parameters. To estimate those uncertainties, we determined 
the errors by perturbing
the observed image intensities by the flux uncertainties 
and then finding a new best fit. In addition, we verified that outside 
the error limits 
for each one of the fitted parameters displayed in Table 1, no satisfying 
solutions were found. To do that, we tried to fit the model with
values outside the parameter bounds specified in the table, and could not find
satisfying solutions in any case. 

An important result is that the derived ejection velocities are almost 
independent on size 
($\gamma$=0.04$\substack{+0.06\\-0.02}$), ranging from 
about 10 to 8 cm s$^{-1}$ for particles
between the lower and upper size limits (130 $\mu$m and 28 cm). This
flat dependence of ejection velocities on size have been previously
inferred in our analysis of the outburst of the asteroid (596)
Scheila, for which we found $\gamma$=0.05 \citep{Moreno11a} . This 
dependence is markedly different to that expected from gas drag by ice 
sublimation processes as occur 
in most comets, with typical values of $\gamma\sim$0.5,
and could be associated to a collision event as was
suggested for Scheila \citep{Moreno11a}, although it cannot be confirmed. 
On the
other hand, if we equate the velocity of the slowest moving 
particles ($v$=8 cm s$^{-1}$ for $r$=28 cm particles) to the escape
velocity, we obtain a nucleus radius of $R_n$=107 m to
$R_n$=152 m, considering bulk densities of 1000 to 500 kg m$^{-3}$. 
At this point, it is interesting to note that computer 
simulations at the catastrophic disruption threshold 
reveal that ejecta velocities depend on target size 
and that ejection velocities of the order of those obtained here 
($\sim$10 cm s$^{-1}$) could be compatible with both porous and non-porous 
targets of $\sim$100 m radius \citep{Jutzi10}. Our derived target radius 
is a far more stringent constraint than that of $R_n$$<$2.1 km
derived by \cite{Stevenson12} based on the non-detection of the
nucleus on images taken by the Wide-Field Infrared
Survey Explorer (WISE). On the other hand, using the 
formalism by \cite{Bowell89}, and assuming a slope parameter of $G$=0.15, and 
a bulk geometric albedo of $p_v$=0.15, the apparent magnitude of a
$R_n$=125 m nucleus at the same phase angle and geocentric and 
heliocentric distances than P/Gibbs would be just above $m_v$=26, which  
represents a negligible contribution to the brightness compared with
that at the optocenter of the trails. This is compatible with our initial 
hypothesis on that the dust cloud brightness dominates the trail 
optocenter. An object of such a small size 
might have a very small rotation period that could 
be under the critical rotation period for fracture so that a rotational 
disruption could also be argued as the cause of the event \citep{Jewitt12},
provided it produces a sudden release of material. A small-sized
nucleus was also derived for P/2010 A2 (LINEAR), the innermost MBC discovered so
far, whose tail has been also reported by many authors as the
result of an impulsive event 
\citep[e.g.][]{Jewitt10,Snodgrass10,Hainaut12}, and not as the result 
of a sustained activity as we suggested \citep{Moreno10}. 
   
The size distribution power index, which essentially controls the
slope of the trail brightness, is --3.7$\pm$0.1. The maximum ejected
particle size derived is $r$=28$\pm$10 cm (Table 1). A 
maximum size smaller than the lower bound would imply a significant displacement
of the maximum of brightness trailward. Conversely, a size 
larger than the upper bound would make the slope of the synthetic 
trail much steeper than observed.

The minimum ejected dust mass is (5$\pm$2)$\times$10$^8$ kg, which is 
near the value reported by \cite{Hainaut12} for P/2010 A2
(8$\times$10$^8$ kg). This is an order of 
magnitude higher than reported by \cite{Stevenson12}, although their
result is just based on a measure of  
the total geometric cross section of the trail 
and assuming a mean size for the ejected particles, not a size
distribution. This total mass is in fact a lower limit, since we
cannot precise neither the amount of small ($r<$100 $\mu$m) particles released,
nor the component of particles traveling at high speed that could result as
consequence of e.g. a collision, if this were the case.          

The best-fit event date (2011 July 1st) agrees with that derived by
\cite{Stevenson12} within the error limits (they derived 2011 July 7th
as the event date). The duration of the event is constrained to be of 
less than a day. In the $\Delta t<$1 day space, it is always possible
to find a set of fitting parameters giving similar values of
$\sigma$. The fits shown in Figure 3 correspond to an event duration
of 0.13 day, i.e., just above 3 hours. An impulsive event is, then,
clearly implied.

\section{Conclusions}  

From the Monte Carlo dust tail modeling of the observations of
Main-Belt Comet P/2012 F5 (Gibbs) we can derive the following conclusions:  

1) The direct study of the MBC images in term of synchrone analysis
imply an impulsive event 
as the cause of the observed trails. We predict the event to have
occurred on 2011 July 1st, with an accuracy of $\pm$10 days. In this 
respect, we agree with the results recently reported by
\cite{Stevenson12} for this MBC, not  
only in the  
nature, but also in the event date within the errors (they report on a
impulsive event on 2011 July 7th with an uncertainty of $\pm$20
days). The duration of the event is constrained to be less than 1
day, possibly less than a few hours. 
   
2) The nature of the impulsive event is impossible to determine with the 
sole information of the physical parameters derived from this 
analysis. Among the likely causes, an outburst, 
a collision with another body, or
a rotational disruption could be invoked. Activity related to  
ice sublimation seems unlikely on the basis of the dependence of the
particle ejection velocities on size, which interestingly 
turns out to be very similar to that we found for (596) Scheila 
\citep{Moreno11a}, a very likely impacted asteroid.   

3) The total dust mass released is $\gtrsim$5$\times$10$^8$ kg. The ejected
particles are distributed in size following a power-law of index
--3.7$\pm$0.1. The maximum particle size ejected is about 30 cm, with a
velocity of $\sim$8 cm s$^{-1}$.  Adopting this value as the escape
velocity, the size of the nucleus is constrained to about 100-150 m in
radius, for bulk body densities in the range 1000-500 kg m$^{-3}$.

\acknowledgments

This article is based on observations made with the Gran Telescopio
Canarias (GTC), installed in the Spanish Observatorio del Roque de los
Muchachos of the Instituto de Astrof\'\i sica de Canarias, in the island 
of La Palma. 

We gratefully acknowledge the comments and suggestions of an anonymous referee.

This work was supported by contracts AYA2009-08190, AYA2011-30613-C02-01, and 
FQM-4555 (Proyecto de Excelencia, Junta de Andaluc\'\i a). 
J. Licandro gratefully acknowledges support from the Spanish ``Ministerio de
Ciencia e Innovaci\'on'' project AYA2011-29489-C03-02.

\clearpage

\begin{figure}[ht]
\centerline{\includegraphics[scale=0.8,angle=-90]{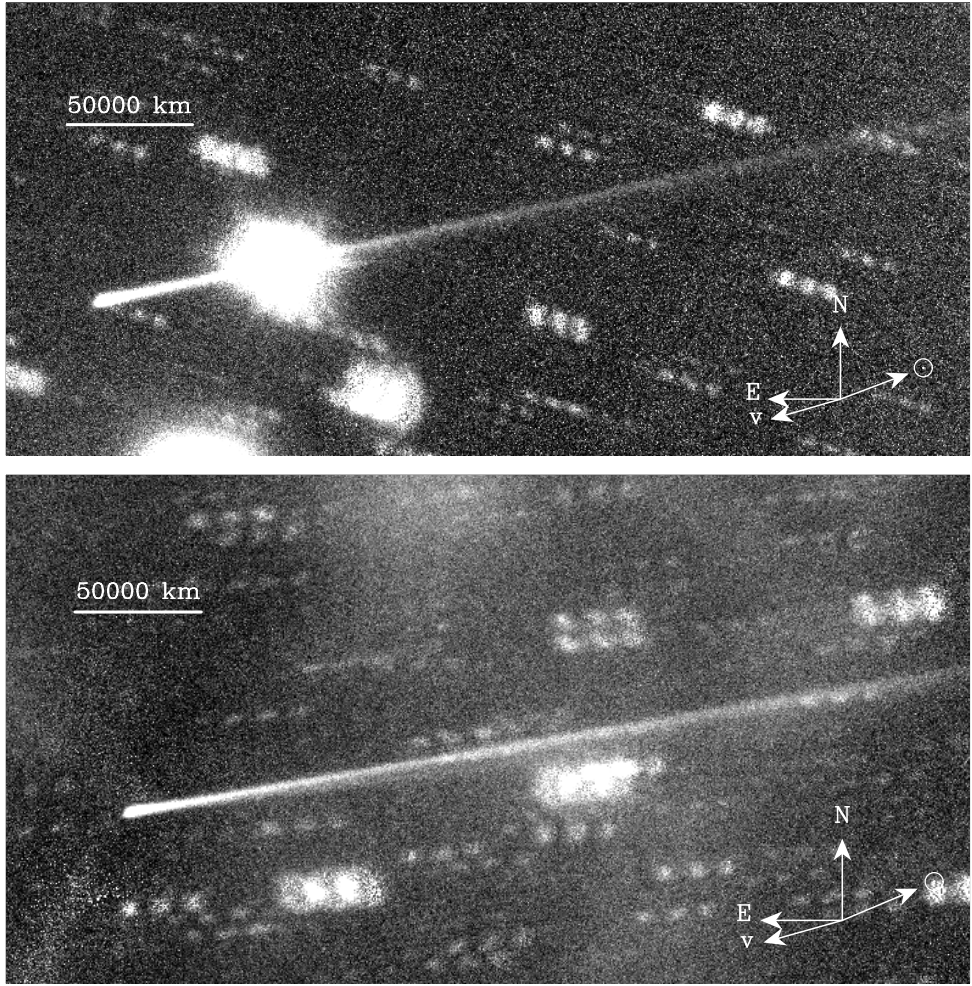}}
\caption{Median stack images of P/2012 F5 (Gibbs) obtained with the OSIRIS
  instrument of the 10.4m Gran Telescopio Canarias through a 
Sloan $r^\prime$ filter, on UT 2012 May 18.9 (upper panel) and June 8.9
(lower panel). The directions of the velocity vector, the Sun, and the
astronomical North and East are indicated. 
   \label{fig1}}
\end{figure}

\clearpage

\begin{figure}[ht]
\centerline{\includegraphics[scale=0.8,angle=-90]{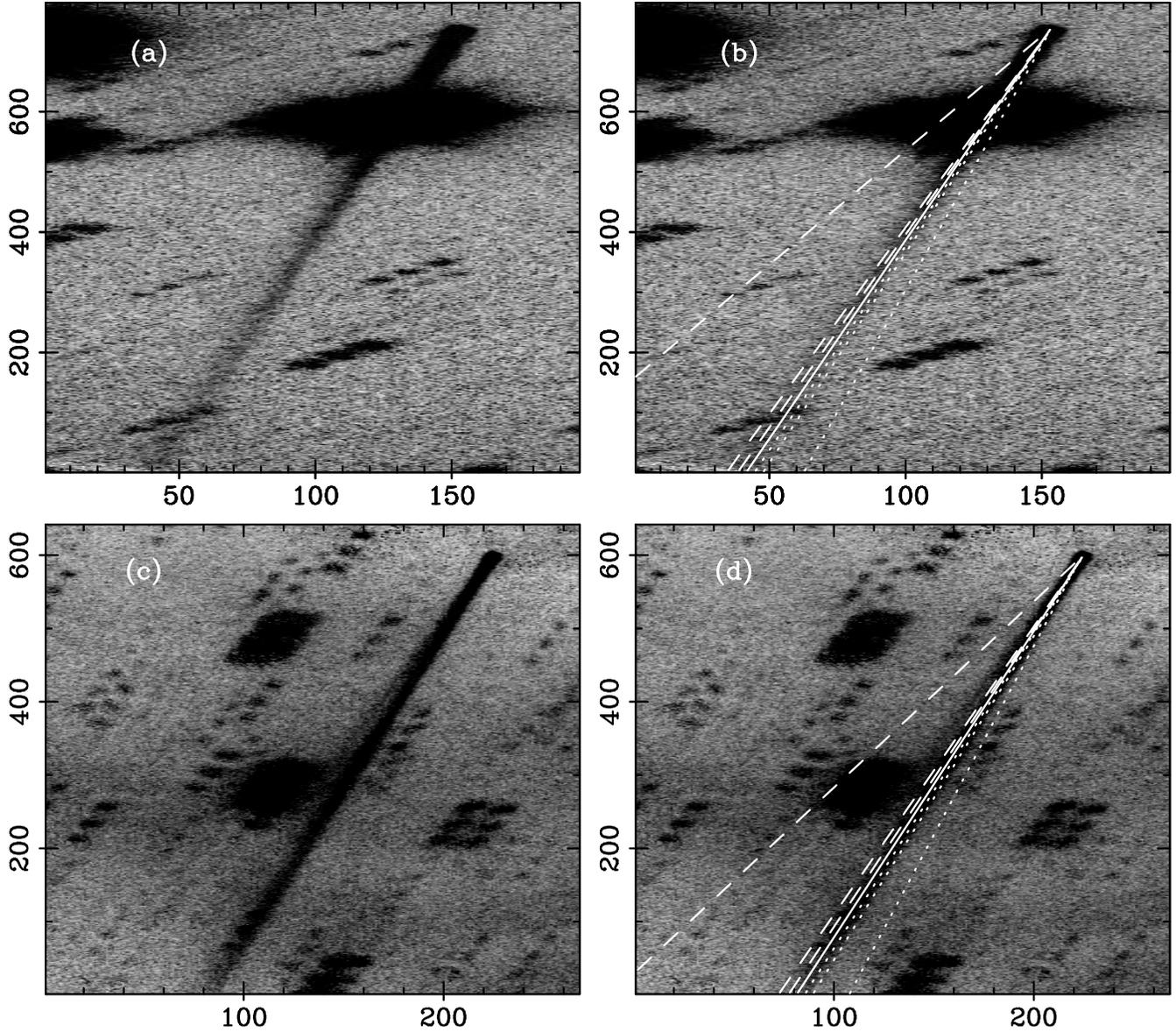}}
\caption{Panel(a) displays the P/2012 F5 (Gibbs) image
  obtained in 2012 May 18.9. Panel (b) shows the same image but with
  seven synchrones overlaid, corresponding, in clockwise order, to
  --100, --20, --10 (dotted lines), 0 (solid line), +10, +20, and +100
  (dashed lines) days respect to the 
  2011 July 3 synchrone. Panels (c) and (d) shows the same as (a) and
  (b), respectively, but for the 2012 June 8.9 image. The scale 
is 485 km px$^{-1}$, and  540 km px$^{-1}$, in the upper and lower
panels, 
respectively. Note that the $x$ and $y$ axes are scaled independently to
facilitate comparison between synchrones.
\label{fig2}}
\end{figure}

\clearpage

\begin{figure}[ht]
\centerline{\includegraphics[scale=0.78,angle=-90]{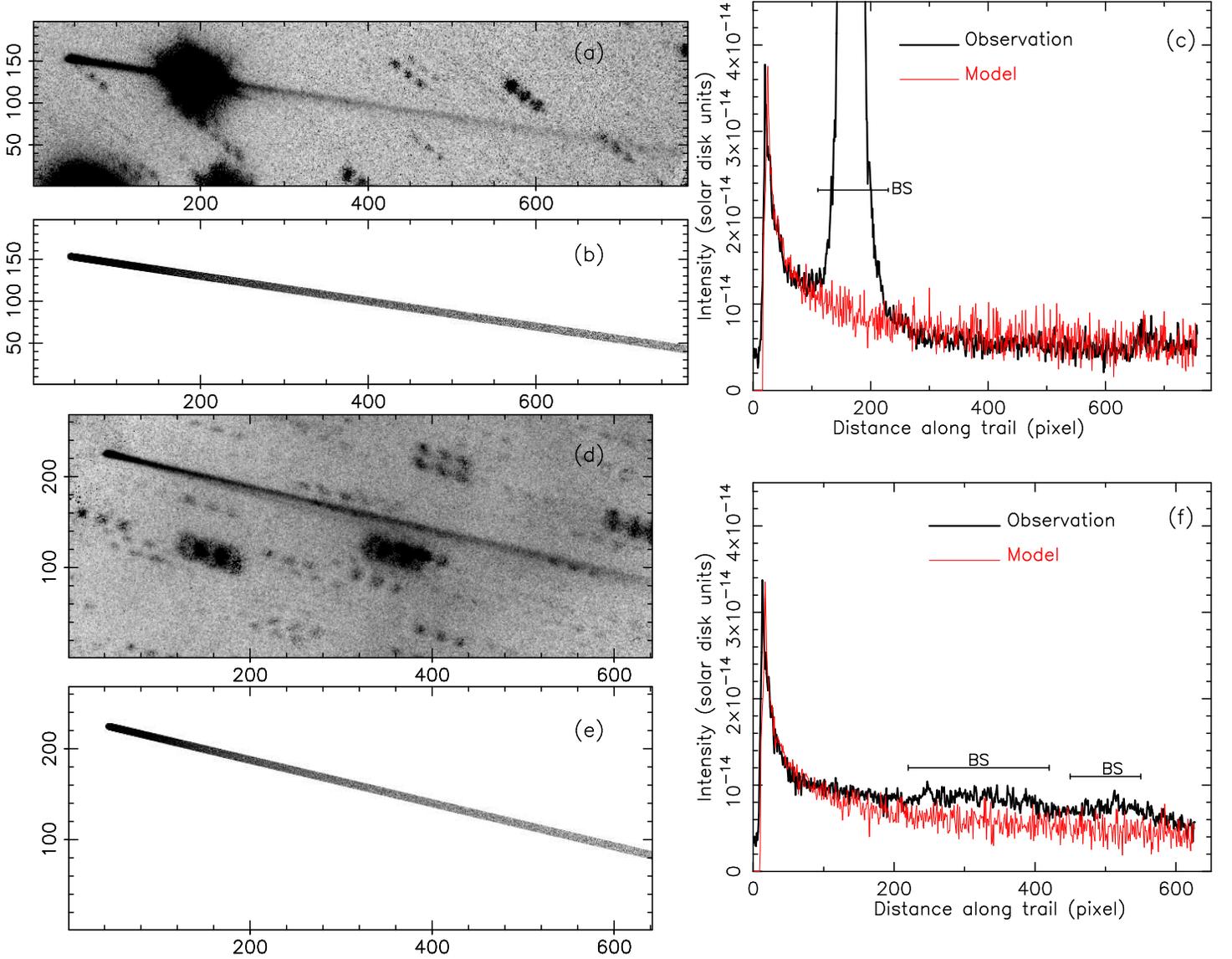}}
\caption{Panels (a) and (b) correspond to the observation and best
  fitted  image for the observation on 2012 May 18th. Panel (c) shows
  observed and modeled scans along the trail of those images. Panels (d), (e), and (f)
  give the same as (a), (b), and (c), respectively, 
but for the 2012 June 8th image. The
  spatial scales are 485 km px$^{-1}$, and  540 km px$^{-1}$,
  respectively. Note
  that the images are rotated by 90 degree counterclockwise so that their $x$ and $y$ axes correspond to
  the photographic $M$ and $N$ axes.  The segments labeled as ``BS''
  pertain to regions of field star contamination in the trails.
\label{fig3}}
\end{figure}

\clearpage

\begin{deluxetable}{ccccccc}
\tablewidth{0pt}
\tablecaption{The best-fit parameters of the model.}
\tablehead{
\colhead{$v_0$} & \colhead{Velocity} & \colhead{SD power} & 
\colhead{Max. radius} & \colhead{Mass lower} &  \colhead{Event}
& Event \\
\colhead{(cm s$^{-1}$)} & \colhead{index ($\gamma$)} & \colhead{index ($\alpha$)}  & 
\colhead{($r_{max}$,cm)} & \colhead{limit($M_e$,kg)} & \colhead{time
  ($t_s$,UT)} & \colhead{duration ($\Delta t$)} \\
}
\startdata
 13$\substack{+5\\-8}$ & 0.04$\substack{+0.06\\-0.02}$ & --3.7$\pm$0.1
 & 28$\pm$10 & (5$\pm$2)$\times$10$^8$ & 2011/07/01$\pm$10d & 0.13d ($<$1d) \\
\enddata
\end{deluxetable}

\begin{thebibliography}{}

\bibitem[Bertini(2011)]{Bertini11} Bertini, I. 2011, Planet. Space
  Sci., 59, 365

\bibitem[Bodewits et al.(2011))]{Bodewits11} Bodewits, D., Kelley,
M.S., Li, J.Y., et al. 2011, \apj, 
 733, L3

\bibitem[Bowell et al.(1989))]{Bowell89}Bowell, E., Hapke, B.,
  Domingue, D. et al., 1989, in Asteroids II, ed. R.P. Binzel,
  T. Gehrels, and M.S. Matthews, Univ. of Arizona Press, Tucson, p. 524.

\bibitem[Cepa et al.(2000))]{Cepa00} Cepa, J., Aguiar, M., Escalera,
  V. et al. 2000, Poc. SPIE, 4008, 623

\bibitem[Cepa(2010))]{Cepa10} Cepa, J. 2010, Highlights of Spanish
  Astrophysics V, Astrophysics and Space Science Proceedings, 
 Springer-Verlag, p. 15

\bibitem[Edoh(1983)]{Edoh83} Edoh, O. 1983, PhD thesis, Univ. Arizona

\bibitem[Finson \& Probstein(1968)]{Finson68} Finson, M., \& Probstein,
  R. 1968, \apj, 154, 327

\bibitem[Fulle(1989)]{Fulle89} Fulle, M., 1989, A\&A, 217, 283

\bibitem[Fulle et al.(2010)]{Fulle10} Fulle, M., Colangeli, L.,
  Agarwal, J., et al. 2010, A\&A, 522, 63

\bibitem[Gibbs et al.(2012)]{Gibbs12} Gibbs, A.R., Sato, H., Ryan,
  W.H. et al. 2012, Central Bureau Electronic Telegrams, 3069, 1

\bibitem[Giblin(1998)]{Giblin98} Giblin, I. 1998, Planet. Space Sci.,
  46, 921

\bibitem[Haghighipour(2009)]{Haghighipour09} Haghighipour, N. 2009 
Meteor. \& Planet. Sci., 44, 1863

\bibitem[Hainaut et al.(2012)]{Hainaut12} Hainaut, O.R., Kleyna, J., 
Sarid, G. et al. 2012, A\&A 537, A69

\bibitem[Hsieh et al.(2004)]{Hsieh04} Hsieh, H.H., Jewitt, D., \& 
Fern\'andez, Y. 2004, \aj, 127, 2997 

\bibitem[Hsieh \& Jewitt(2006)]{Hsieh06} Hsieh, H.H., \& Jewitt,
  D. 2006, Science, 312, 561

\bibitem[Hsieh et al.(2009)]{Hsieh09} Hsieh, H.H., \& Jewitt, D., \&
  Ishiguro, M. 2009, \aj, 137, 157

\bibitem[Hsieh et al.(2010)]{Hsieh10} Hsieh, H.H., \& Jewitt, D., 
  Lacerda, P. 2010, \mnras, 403, 363

\bibitem[Hsieh et al.(2011)]{Hsieh11} Hsieh, H.H., Meech, K., \&
  Pittichova, J.  2011b, \apj, 736, L18

\bibitem[Hsieh et al.(2012a)]{Hsieh12a} Hsieh, H.H., Yang, B., \&
  Haghighipour, N. 2012, \apj, 744, 9

\bibitem[Hsieh et al.(2012b)]{Hsieh12b} Hsieh, H.H., Yang, B., 
  Haghighipour, N., et al. 2012, \apj, 748, L15

\bibitem[Hsieh et al.(2012c)]{Hsieh12c} Hsieh, H.H., Yang, B., 
  Haghighipour, N., et al. 2012, \aj, 143, 104

\bibitem[Ipatov \& Hahn(1999)]{IpatovHahn99} Ipatov, S.I., \& Hahn, G.J.
  1999, Solar Syst. Res. , 33, 487

\bibitem[Ishiguro et al.(2011)]{Ishiguro11} Ishiguro, M., Hanayama,
  H., Hasegawa, S. et al. 2011, \apj, 740, L11

\bibitem[Jewitt et al.(2009)]{Jewitt09} Jewitt, D., Yang, B., \&
  Haghighipour, N. 2009 \aj, 137, 4313

\bibitem[Jewitt et al.(2010)]{Jewitt10} Jewitt, D., Weaver, H.,
  Agarwal, J., Mutchler, M.,  \& Drahus, M. 2010 Nature, 467, 817




\bibitem[Jewitt et al.(2011)]{Jewitt11} Jewitt, D., Weaver, H., 
  Mutchler, M., et al. 2011 \apj, 733, L4

\bibitem[Jewitt(2012)]{Jewitt12} Jewitt, D. 2012 \aj, 143, 21

\bibitem[Jutzi et al.(2010)]{Jutzi10} Jutzi, M., Michel, P., Benz, M., \& 
Richardson, B.C. 2010 Icarus, 207, 54

\bibitem[Licandro et al.(2011)]{Licandro11} Licandro, J., Campins, H., 
Tozzi, G.P., et al. 2011 A\&A 532, 65

\bibitem[Licandro et al.(2012)]{Licandro12} Licandro, J., de Le\'on,
  J., Moreno, F., et al. 2012 A\&A, submitted

\bibitem[Moreno(2009)]{Moreno09} Moreno, F. 2009, \apjs, 183, 33

\bibitem[Moreno et al.(2010)]{Moreno10} Moreno, F., Licandro, J.,
 Tozzi, G.-P., et al. 2010, \apj, 718, L132

\bibitem[Moreno et al.(2011a)]{Moreno11a} Moreno, F., Licandro, J.,
 Ortiz, J.L., et al. 2011a, \apj, 738, 130

\bibitem[Moreno et al.(2011b)]{Moreno11b} Moreno, F., Lara, L.M., 
Licandro, J., et al. 2011b, \apj, 738 L16

\bibitem[Moreno et al.(2012)]{Moreno12} Moreno, F., Pozuelos, F.,
  Aceituno, F., et al. 2012, \apj, 752, 136

\bibitem[Nelder \& Mead(1965)]{Nelder65} Nelder, J. A., \& 
Mead, R. 1965, Comput. J., 7, 308

\bibitem[Onose \& Fujiwara(2004)]{Onose04} Onose, N., \& Fujiwara,
  A. 2004, Meteoritics and Planet. Sci., 39, 321

\bibitem[Press et al.(1992)]{Press92} Press, W.H., 
Teukolsky, S.A., Vetterling, W.T., \& Flannery, B.P. 1992, in
Numerical Recipes in FORTRAN (Cambridge: Cambridge Univ. Press), 402

\bibitem[Snodgrass et al.(2010)]{Snodgrass10} Snodgrass, C., Tubiana,
  C., Vincent, J.-B. et al. 2010 Nature, 467, 814

\bibitem[Stevenson et al.(2012)]{Stevenson12} Stevenson, R., Kramer,
  E.A., Bauer, J.M. et al. 2012, \apj, in press

\bibitem[Yang \& Hsieh(2011)]{YangHsieh11} Yang, B.,  \& Hsieh, H.H.,
  2011, \apj, 737, L39


\end{thebibliography}
\end{document}